\documentclass[pra,aps,twocolumn,showpacs,epsfig]{revtex4}
\newcommand{\beq}{\begin{equation}}
\newcommand{\eeq}{\end{equation}}
\newcommand{\beqa}{\begin{eqnarray}}
\newcommand{\eeqa}{\end{eqnarray}}
\newcommand{\ba}{\begin{array}}
\newcommand{\ea}{\end{array}}
\usepackage{color}
\usepackage[dvips]{epsfig}

\begin{document}

\title{Spontaneous symmetry breaking and collapse \\ 
in bosonic Josephson junctions} 
\author{Giovanni Mazzarella$^{1}$ and Luca Salasnich$^{1,2,3}$}
\affiliation{$^{1}$Dipartimento di Fisica ``Galileo Galilei'' and 
Consorzio Nazionale Interuniversitatio per le Scienze Fisiche 
della Materia (CNISM), 
Universit\`a di Padova, Via Marzolo 8, I-35122 Padova, Italy \\
$^{2}$Istituto Nazionale di Ottica (INO) del Consiglio Nazionale 
delle Ricerche (CNR), Research Unit at Padova, 
Via Marzolo 8, I-35122 Padova, Italy\\
$^{3}$Center for Applied Mathematics and Theoretical Physics, 
University of Maribor, Krekova 2, SI-2000 Maribor, Slovenia}

\begin{abstract}
We investigate an attractive atomic Bose-Einstein 
condensate (BEC) trapped by a double-well potential in the axial 
direction and by a harmonic potential in the transverse 
directions. We obtain numerically, for the first time, 
a quantum phase diagram which includes all 
the three relevant phases of the system: 
Josephson, spontaneous symmetry breaking (SSB), and 
collapse. We consider also the coherent dynamics of the BEC and 
calculate the frequency of population-imbalance mode 
in the Josephson phase and in the SSB phase up to the collapse. 
We show that these phases can be observed by using ultracold
vapors of $^7$Li atoms in a magneto-optical trap.
\end{abstract}


\maketitle

In many experiments atomic Bose-Einstein condensates 
(BECs) are cigar-shaped due to a strong harmonic 
trapping potential in the cylindric radial plane; 
these BECs can be separated in two parts by means of a 
double-well potential in the cylindric axial direction \cite{oliver}. 
This kind of geometry is the ideal setup to 
study the Josephson effect, a macroscopic coherent
phenomenon which has been observed in systems as diverse as
superconductors \cite{book-barone},
superfluid Helium \cite{helium} and, recently, also
BECs in trapped ultracold atomic gases \cite{exp-bec}.
The observed coherent dynamics of the atomic BEC in the double-well
potential (bosonic Josephson junction) \cite{oliver,exp-bec}
is efficiently described by nonlinear Josephson equations (JEs)
\cite{smerzi}, which are based on a two-mode approximation
of the Gross-Pitaevskii equation (GPE) \cite{leggett}. 
These JEs are fully symmetric by changing the 
sign of the inter-atomic scattering length and do not predict 
the collapse of the BEC. The collapse of an attractive BEC of $^7$Li atoms 
or $^{85}$Rb atoms  has been observed by various 
experimental groups \cite{col-exp}   
and theoretically analyzed by many authors: 
in a single-well potential \cite{col-single}, in a potential 
without axial confinement \cite{col-soliton}, 
in a toroidal confinement \cite{col-torus}, 
in a double-well potential \cite{col-double}, 
and in a periodic potential \cite{col-periodic}. 

In this paper, by correctly taking into account the dimensional 
reduction of GPE from 3D to 1D, i.e. by using the so-called 
1D nonpolynomial Schr\"odinger equation (1D NPSE) \cite{salasnich}, 
we show that for an attractive BEC (negative inter-atomic 
scattering length) the JEs are not reliable in the presence of 
strong coupling. By numerically solving the 1D NPSE we obtain, 
for the first time, a quantum phase diagram of the three 
relevant regimes of the attractive BEC in a double-well: 
the Josephson phase, where the metastable state of lowest finite energy 
has a balanced population \cite{book-barone}, 
the spontaneous symmetry breaking (SSB) phase, 
where the metastable state has an unbalanced population 
\cite{oliver,smerzi}, and the phase of collapse, where the system reaches the 
collapsed ground-state with energy equal to minus infinity. 
Note that the problem of BEC collapse in an axial double well 
potential has been investigated by Sakellari, Proukakis, and Adams 
\cite{col-double}, but they have not derived the quantum phase diagram 
of the attractive BEC. Instead, very recently the collapse region in 
a quantum phase diagram has been obtained for 
a pair of cigar-shaped traps coupled by 
tunneling of atoms \cite{io-e-boris}. 
We also study the coherent dynamics of the system and calculate the 
frequency of the population imbalance both in the Josephson regime  
and in the SSB regime. In the SSB phase this frequency 
reaches its maximum value at the coupling strength where there 
is the collapse of the BEC. In addition, from the 1D NPSE we obtain  
generalized Josephson equations, which we call 
nonpolynomial Josephson equations (NPJEs) for the fractional 
imbalance and relative phase of the bosonic Josephson junction.
These new NPJEs reduce to the familiar JEs 
in the weak-coupling limit, but show a better agreement with 
the numerical results of the 1D NPSE (and 3D GPE) for strong couplings 
(for both positive and negative scattering length). 
Finally, we suggest that our predictions can be observed experimentally
by using an ultracold vapor of $^7$Li atoms and tuning the
$s$-wave scattering length.

Let us consider a dilute interacting BEC at zero temperature confined
by a trapping potential $V_{trap}({\bf r})$. This 
potential is taken to be the superposition of an isotropic harmonic
confinement in the the transverse radial plane
and a double-well potential $V_{DW}(x)$ in the axial direction $x$.
Then, $V_{trap}({\bf r})$ is given by
\beq 
V_{trap}({\bf r})= V_{DW}(x) +\frac{m\omega_{\bot}^2}{2}\rho^2 \; , 
\eeq 
where $\rho$ is the cylindric radial coordinate,
$m$ is the mass of the atom, and $\omega_{\bot}$ is the trapping
frequency in the radial plane. The macroscopic wave function
$\Psi({\bf r},t)$ describing the above system with $N$ atoms
is governed by the 3D GPE 
\beq 
i \hbar {\partial \over \partial t} \Psi = 
\left[ -{\hbar^2\over 2m} \nabla^2 + V_{trap}({\bf r}) 
+ {4 \pi \hbar^2 a_s N \over m} |\Psi|^2 \right] \Psi 
\label{3dgpe}
\eeq
where $a_s$ the $s$-wave boson-boson scattering
length and $\Psi({\bf r},t)$ is normalized to $1$. 
The 3D GPE captures the main properties of collapse threshold and, 
as shown by using a reliable nonlocal potential, 
the collapsed state is actually a state of very high density which 
decays due to inelastic two- and 
three-body collisions \cite{salasnich-old}. 
By following Ref. \cite{salasnich}, we choose the wave function
$\Psi({\bf r},t)$ as the product of an axial complex wave
function $f(x,t)$ and a Gaussian transverse wave function of
radial width $\sigma$, where $\sigma$ depends
on the axial wave function $f(x,t)$, i.e. $\sigma=\sigma(f(x,t))$.
By expressing lengths in units of $a_{\bot}=\sqrt{\hbar/m
\omega_{\bot}}$, times in units of $\omega_{\bot}^{-1}$, and energies in units
of $\hbar \omega_{\bot}$, it is easy to show that the fields $f(x,t)$ and
$\sigma(x,t)$ satisfy the following equations \cite{salasnich}
\beq
i\frac{\partial f}{\partial t}=\bigg[-\frac{1}{2}
\frac{\partial ^2}{\partial x^2}+V_{DW}(x)
+ \frac{1}{2}\bigg(\frac{1}{\sigma^2}+\sigma^2\bigg)
+ \frac{\Gamma|f|^2 }{\sigma^2}\bigg]f \; ,
\label{el1}
\eeq
\beq
\sigma^4=1+\Gamma|f|^2 \; ,
\label{el1b}
\eeq
where $\Gamma=2a_s N/a_{\bot}$ and $f(x,t)$ is normalized to $1$.
Inserting Eq. (\ref{el1b}) into Eq. (\ref{el1}) one gets the
so-called 1D NPSE \cite{salasnich}, which is extremely
accurate in reproducing the properties of the full 3D GPE 
with transverse harmonic confinement \cite{salasnich}. 

\begin{figure}[ht]
\centering
\epsfig{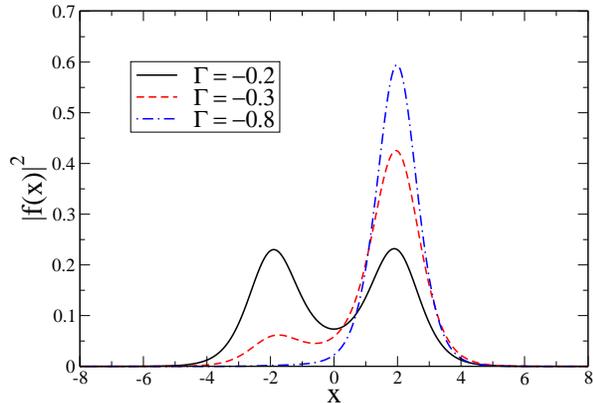}
\caption{(color online). Axial probability density $|f(x)|^2$
of the metastable attractive BEC in the symmetric double-well 
potential $V_{DW}(x)$  where the two minima are at $x=\pm x_0$ 
with $x_0=2$ and the energy height is $U_0=0.8$. $\Gamma=2Na_s/a_{\bot}$ 
is the interaction strength. Results obtained by using 1D NPSE, 
Eqs. (\ref{el1}) and (\ref{el1b}). 
Length $x$  in units of $a_{\bot}=\sqrt{\hbar/m\omega_{\bot}}$, 
density $|f|^2$ in units of $a_{\bot}^{-1}$ and energy 
in units of $\hbar \omega_{\bot}$.} 
\label{fig1}
\end{figure}

We have solved the 1D NPSE by using a finite-difference 
Crank-Nicolson code with imaginary time \cite{salasnich2,sala-numerics} 
to obtain the ground-state of BEC in the symmetric double-well trap.  
In the numerical analysis the double-well potential
$V_{DW}(x)$ is given by the combination of two P\"{o}schl-Teller
potentials with the energy barrier of height $U_0=0.8 \, \hbar \omega_{\bot}$
and the local minima  at $-x_0=-2a_{\bot}$ and $x_0=2a_{\bot}$
(for details see Refs. \cite{mazza,mazza2}). 
It is important to stress that with $\Gamma <0$ the ground-state 
is always the collapsed state with energy equal to minus infinity. 
Thus, for $\Gamma <0$ we are actually looking for the metastable 
state of lowest finite energy. We find that this metastable state is  
symmetric for $\Gamma_{SSB}<\Gamma < 0$ (Josephson phase), 
it has a broken symmetry for $\Gamma_C < \Gamma < \Gamma_{SSB}$ 
(SSB phase), and it becomes the collapsed ground-state 
for $\Gamma < \Gamma_C$ (collapsed phase). 
In Fig. \ref{fig1} we plot the axial probability density $|f(x)|^2$ 
of the metastable state 
obtained by solving the 1D NPSE with imaginary time. 
We start with a slightly asymmetric initial condition 
and proceed up to the convergence to a stable configuration. 
The figure shows that for $\Gamma = -0.2$ the profile of the (meta-)stable 
state is symmetric, while for $\Gamma = -0.3$ it is not. 
For $\Gamma = -0.8$ the 
BEC is practically localized only in the right well. In addition, 
we find that for $\Gamma < -1.2$ there is the collapse. 

\begin{figure}[ht]
\centering
\epsfig{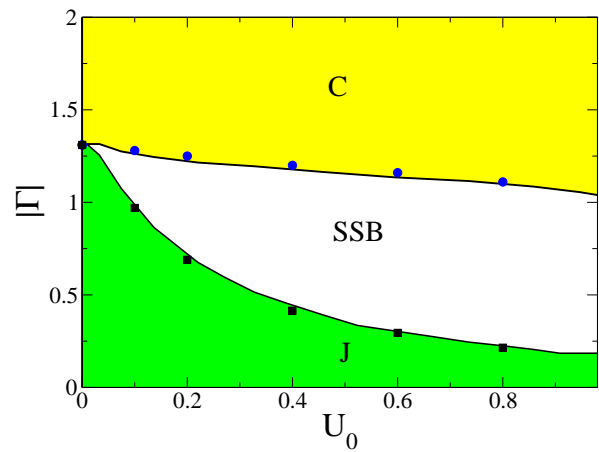}
\caption{(color online). Quantum phase diagram ($U_0,|\Gamma|$) 
of the attractive BEC in the symmetric double-well 
potential $V_{DW}(x)$. The two minima are at $x=\pm x_0$ 
with $x_0=2$. $U_0$ is height the of the central barrier of $V_{DW}(x)$ 
and $\Gamma=2Na_s/a_{\bot}$ is the interaction strength. 
There are three phases: Josephson (J), spontaneous symmetry breaking 
(SSB), and collapse (C). Solid lines are obtained with 1D NPSE, 
Eqs. (\ref{el1}) and (\ref{el1b}). Filled circles and squares 
are obtained with 3D GPE, Eq. (\ref{3dgpe}). Energy $U_0$ in units of
$\hbar\omega_{\bot}$.}
\label{fig2}
\end{figure}

It is very interesting to analyze the quantum phases of the attractive BEC 
as a function of the height $U_0$ of the energy barrier. 
We have performed a systematic 
investigation by changing both $U_0$ and $\Gamma$. The results are 
shown in Fig. \ref{fig2} where we plot the quantum phase 
diagram in the plane ($U_0,|\Gamma|$). 
To our knowledge, this is the first time 
that this kind of phase diagram is obtained for an 
attractive BEC in a symmetric double-well potential. 
In the figure, the solid lines are obtained from the 1D NPSE, 
Eqs. (\ref{el1}) and (\ref{el1b}). The interaction 
strength $|\Gamma_{SSB}|$ of the SSB transition strongly 
decreases by increasing the height $U_0$ of the energy barrier, 
while the critical stregth $|\Gamma_C|$ to get the collapse 
slightly decreases by increasing $U_0$. 
To check the accuracy of the NPSE we have also solved the 3D GPE, 
Eq. (\ref{3dgpe}), by using a cylindric-symmetry finite-difference 
Crank-Nicolson code with imaginary time \cite{sala-numerics}. 
In Fig. \ref{fig2} the collapse points predicted by 3D GPE 
are shown as filled circles, while the SSB points of 3D GPE 
are filled squares. As expected \cite{salasnich}, the agreement 
between 1D NPSE and 3D GPE is quite good. 
We stress, however, that both 3D GPE and 1D NPSE are based 
on a zero-range inter-atomic potential \cite{leggett,salasnich}. 
A more accurate description of interaction might lead to a slightly 
different transition between the SSB and the collapse 
phase in Fig. \ref{fig2}. 

Let us now consider this two-mode approximation of the 1D NPSE. 
Under the condition that the central barrier $U_0$ of the double-well 
potential $V_{DW}(x)$ is sufficiently high -
corresponding to a weak link between its left ad right
sides - the field $f(x,t)$ can be decomposed by using the two-mode
approximation
\beq
f(x,t)=f_{L}(t)\phi_{L}(x)+f_{R}(t)\phi_{R}(x) \; .
\label{twomode}
\eeq
The functions $\phi_{L}(x)$ and $\phi_{R}(x)$, which are orthonormal,
are localized in the left and right well, respectively.
We assume that the above functions are real and use
the ansatz (\ref{twomode}) in the NPSE. We 
multiply the resulting equation by 
$\phi_{\alpha}(x)$ ($\alpha=L,R$), and integrate over $x$.
Then, by taking into account the overlaps 
between $\phi_{\alpha}$'s localized in the same
well and neglecting those ones between $\phi_{\alpha}$'s localized 
in different wells, we obtain
\beqa
i\frac{\partial f_{\alpha}}{\partial t} &=&
\bigg[\frac{1}{2}\bigg(\frac{1}{\sigma_{\alpha}^{2}}+
\sigma_{\alpha}^2\bigg)
+ \frac{g}{\sigma_{\alpha}^2}|f_{\alpha}|^2 
+\epsilon \bigg]f_{\alpha}-K f_{\beta} \; ,
\label{el2}
\eeqa
\beq
\sigma_{\alpha}^4=1+g|f_{\alpha}|^2 \; ,
\label{el2b}
\eeq
where the parameters $\epsilon$, $K$ and $g$ are given by
\beqa
\label{parameters}
\epsilon &=& \int_{-\infty}^{+\infty} dx \phi_{\alpha}(x)\bigg[-\frac{1}{2}
\frac{\partial
^2}{\partial x^2}+V_{DW}(x)\bigg]\phi_{\alpha}(x)
\\
K &=& \int_{-\infty}^{+\infty} dx\phi_{\beta}(x)
\bigg[-\frac{1}{2}\frac{\partial
^2}{\partial x^2}+V_{DW}(x)\bigg]\phi_{\alpha}(x)
\\
g &=& \Gamma \int_{-\infty}^{+\infty} dx (\phi_{\alpha}(x))^4 \; . 
\eeqa
Finally, we write the time-dependent amplitudes $f_{\alpha}(t)$ as
$
f_{\alpha}(t)=\sqrt{N_{\alpha}(t)}\,e^{i\theta_{\alpha}(t)} \
$,
with $N_{\alpha}(t)$ being the fraction of bosons in the $\alpha$-th well
and $\theta_{\alpha}(t)$ the corresponding phase. Then, 
Eqs. (\ref{el2}) and (\ref{el2b}) give rise to the following 
system of coupled ordinary differential 
equations for the fractional imbalance $z(t)=N_L(t)-N_R(t)$
(here $N_L(t)+N_R(t)=1$) and the relative phase
$\theta(t)=\theta_R(t)-\theta_L(t)$:
\beqa
\dot{z}&=& -2K\sqrt{1-z^2}\sin\theta \; , 
\label{evolution1}
\eeqa
\beqa
\dot{\theta} &=&2K\frac{z}{\sqrt{1-z^2}}\cos\theta
+\frac{\sqrt{1+z}\big(4+3g(1+z)\big)}
{2\sqrt{2}\sqrt{1+z}\sqrt{2+g(1+z)}}
\nonumber
\\
&-&\frac{\sqrt{1-z}\big(4+3g(1-z)\big)}
{2\sqrt{2}\sqrt{1-z}\sqrt{2+g(1-z)}} \; .
\label{evolution1b}
\eeqa
We call these equations 
"nonpolynomial Josephson equations" (NPJEs) because they are derived 
from the NPSE. NPJEs describe the dynamics of the fractional imbalance
$z(t)$ and relative phase $\theta(t)$ of the bosonic Josephson junction
taking into account transverse-size effects. It is clearly much easier 
to solve numerically these NPJEs than the full 3D GPE or the 1D NPSE.  
When the coupling strength $g$ is much smaller than one 
our NPJEs become
\beqa
\label{evolution3}
&&\dot{z}= -2\,K\,\sqrt{1-z^2}\sin\theta
\\
&&\dot{\theta} =2\,K\,\frac{z}{\sqrt{1-z^2}}\cos \theta + g z \; ,
\label{evolution3b}
\eeqa
which are the familiar Josephson equations (JEs) for a BEC
found in Ref. \cite{smerzi}. 
It is straightforward to verify that Eqs. (\ref{evolution3}) and
(\ref{evolution3b}) are invariant under the transformations
$\Gamma \rightarrow -\Gamma$ and $\theta
\rightarrow - \theta + \pi$. Instead, Eqs. (\ref{evolution1})
and (\ref{evolution1b}) do not exhibit this invariance. 

The stationary Josephson regime corresponds to the 
equilibrium points with $z=0$ and $\theta=0$ (balanced population). 
Both JEs and NPJEs show that this stationary Josephson 
phase exists only for $\Gamma_{SSB} < \Gamma$, 
where $\Gamma_{SSB} <0$. We have verified that the value of 
$\Gamma_{SSB}$ predicted by JEs 
is always very close to the 
value obtained with NPJEs. The points below 
$\Gamma_{SSB}$ correspond to SSB phase. This phase, according 
to the JEs, exists for any $\Gamma_{SSB} < \Gamma$; 
while, according to the NPJEs, the SSB phase 
does not exist anymore at the collapse strength $\Gamma_C$. 
Thus NPJEs predict a collapsed phase while JEs do not. 

\begin{figure}[ht]
\centering
\epsfig{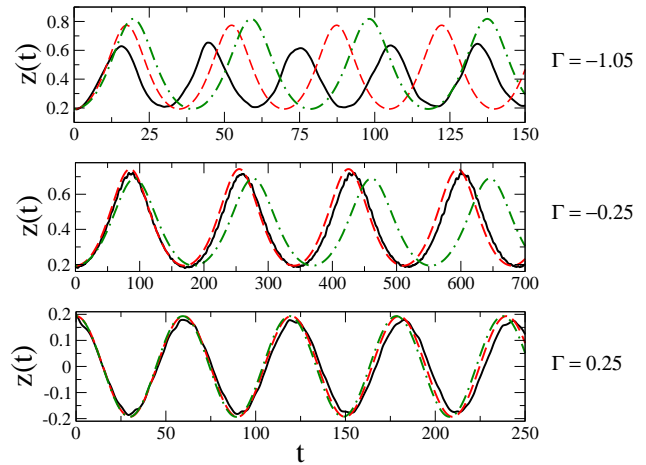}
\caption{(color online). Fractional imbalance $z$ as a function of the time.
Upper panel: $\Gamma =-1.05$.
Middle panel: $\Gamma=-0.25 $. Lower panel: $\Gamma=0.25$.
The solid lines are obtained with 1D NPSE, 
Eqs. (\ref{el1}) and (\ref{el1b}). The dashed lines are obtained with 
NPJEs, Eqs. (\ref{evolution1}) and (\ref{evolution1b}). The dot-dashed line
are obtained with JEs, Eqs. (\ref{evolution3}) and (\ref{evolution3b}). 
Initial conditions: $z(0)=0.2$ and $\theta(0)=0$.
Parameters of the double-well potential $V_{DW}(x)$:
energy barrier height $U_0=0.86$ and location of the two minima 
at $x=\pm x_0$ with $x_0=2$.
Lengths in units of $a_{\bot}$, time in units of
$\omega_{\bot}^{-1}$, energies in units of $\hbar \omega_{\bot}$.}
\label{fig3}
\end{figure}

To compare NPJEs with JEs we plot in Fig. \ref{fig3} the population 
imbalance $z(t)$ for three values of $\Gamma$, choosing as 
initial conditions $z(0)=0.2$ and $\theta(0)=0$. 
The figure shows that the NPJE curves (dashed lines) 
are always closer to the NPSE results (solid lines) than the 
JE ones (dot-dashed lines). Nevertheless, for a sufficiently strong 
(and negative) $\Gamma$ the predictions NPJEs are no longer reliable. 
Notice that while for $\Gamma =0.25$ 
the system displays Josephson oscillations, i.e. coherent oscillations 
around $z=0$, for $\Gamma =0.25$ and 
$\Gamma=-1.05$ there are SSB oscillations, i.e 
coherent oscillations around $z=z_{SSB}\neq 0$. 

\begin{figure}[ht]
\centering
\epsfig{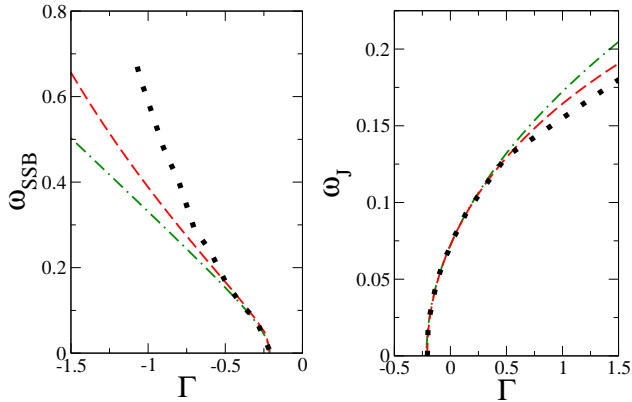}
\caption{(color online). Left panel: spontaneous symmetry breaking 
oscillation frequency $\omega_{SSB}$ 
around $(z=z_{SSB},\theta=0)$ vs. $\Gamma$.
Right panel: Josephson oscillation frequency $\omega_J$
around $(z=0,\theta=0)$ vs. $\Gamma$.
Filled squares are obtained with 1D NPSE, 
Eqs. (\ref{el1}) and (\ref{el1b}). The dashed lines are obtained with 
NPJEs, Eqs. (\ref{evolution1}) and (\ref{evolution1b}). The dot-dashed line
are obtained with JEs, Eqs. (\ref{evolution3}) and (\ref{evolution3b}). 
Parameters of the double well and units as in Fig. \ref{fig3}.}
\label{fig4}
\end{figure}

We investigate in detail these coherent oscillations by 
looking for the stationary points and calculating 
the frequency of small oscillations around these points. 
In the case of NPJEs and JEs we diagonalize 
the Jacobian matrices associated, respectively, to  
the NPJEs (\ref{evolution3})-(\ref{evolution3b}) 
and JEs (\ref{evolution1})-(\ref{evolution1b}). In general, 
the $2\times 2$ Jacobian matrix has two complex eigenvalues, 
$\lambda_{1,2}$, and the stationary point is stable when 
$\lambda_{1,2}=\mp i\omega$ with $\omega >0$ 
the frequency of stable oscillations. 

We consider first the Josephson regime and thus we study 
oscillations around the equilibrium points with $z=0$ 
and $\theta=0$ (balanced population). 
Notice that for $\Gamma =0$ the oscillation frequency $\omega_J$
reduces to the Rabi frequency, i.e. $\omega_J = 2K/\hbar$.
In the right panel of Fig. \ref{fig4} we report the 
frequency $\omega_J$ of coherent oscillations 
around $z=0$ as a function of $\Gamma$. 
From the plots of Fig. \ref{fig4} 
one can see the differences between the behavior of
$\omega_J$ predicted by 1D NPSE (dots), 
NPJEs (dashed line), and JEs (dot-dashed lines). 
Among the three sets of data obviously NPSE ones are the more reliable. 
The softening of $\omega_J$ as $\Gamma<0$ approaches $\Gamma_{SSB}=-0.21$ 
is reproduced extremely well by both NPJEs and JEs, 
while there are differences in corresponence of the hardening 
of $\omega_J$ for large positive values of $\Gamma$. 
At $\Gamma =1.5$ the relative error in the determination of $\omega_{J}$ 
between NPSE and NPJEs results is about $15\%$. 

We now analyze the oscillations in the SSB regime. 
As previously stressed to this regime is associated 
a symmetry breaking of the fractional population 
imbalance $z$, i.e the stationary configuration has $\theta=0$ 
but $z=z_{SSB}\neq 0$. 
In the left panel of Fig. \ref{fig4}, 
we plot the SSB oscillation 
frequency $\omega_{SSB}$ around the stationary $z_{SSB}\neq 0$ 
(with $\theta(0)=0$) as a function of $\Gamma$. From these plots one can see
that when $\Gamma$ is small enough 
the behavior of $\omega_{SSB}$ predicted by
NPJEs and JEs are quite similar, but NPJEs results are slightly better 
than JEs ones. $\omega_{SSB}$ is equal 
to zero at $\Gamma_{SSB}$ and it increases by decreasing $\Gamma <0$. 
As $\Gamma$ approaches the collapse strength $\Gamma_C=-1.1$ 
of 1D NPSE the relative error in the determination of $\omega_{SSB}$ 
between NPSE and NPJEs results becomes quite large: 
it is about $50\%$ at $\Gamma = -1.1$. 
We have verified that, in contrast with JEs, 
NPJEs predict the BEC collapse but 
at a critical strength much smaller (in modulus much larger) 
than the one obtained by using the NPSE. 

It is important to stress that it is possible to achieve 
the so-called ``self-trapping regime'' in correspondence of initial 
conditions $(z(0),\theta(0))$ 
which are not stationary points of JEs \cite{smerzi}. This regime
is characterized by population imbalance ($z(t)\neq 0$) 
and a running phase during the time evolution. By solving our NPJEs 
we find this dynamical self-trapping for 
$\Gamma_c < \Gamma < \Gamma_{(ST,-)}<0$ 
or $\Gamma > \Gamma_{(ST,+)}>0$ with the thresholds
$\Gamma_{(ST,\mp)}$ depending on the initial conditions 
$z(0)$ and $\theta(0)$. Note that by solving JEs one 
finds $\Gamma_c=0$ and also $\Gamma_{ST,-}=-\Gamma_{ST,+}$. 

In conclusion, we observe that the results obtained so far,
can be used to describe concrete
systems. For instance, by considering an
attractive Bose-Einstein condensate made of $^{7}$Li atoms,
and choosing the transverse confining frequency as
$\omega_{\bot } \simeq 2\pi \times 100$ Hz, we have a typical value of the
transverse length $a_{\bot}\simeq 4$ $\mu$m, while the
parameters of the double-well potential read:
$U_0\simeq 6 \cdot 10^{-32}J$ and $x_0\simeq 8$ $\mu$m \cite{oliver,gati}.
The natural scattering length of $^7$Li atoms is $a_{s}=-1.45$ nm
but it can be modified with an external constant magnetic field
by means of a Feshbach resonance \cite{col-exp}. 
Working with $N\simeq 10^4$ condensed atoms in the trap,
it is possible to observe experimentally the behavior 
of Josephson frequency $\omega_J$ and of the SSB 
frequency $\omega_{SSB}$ by tuning the scattering length $a_s$ from 
positive values to the collapse point at 
$a_s=\Gamma_c a_{\bot}/(2N)\simeq -0.2$ nm. 
Finally, we stress that with the above values the 
condition $|g|/K\ll N^2$ is fully satisfied, 
and the system is always in the 
coherent regime \cite{leggett,mazza,mazza2}. 

The authors thank Flavio Toigo for enlightening discussions.

\end{document}